\documentstyle[eqsecnum,prl,epsfig,preprint,aps]{revtex}
\begin{document}
\tightenlines
\bibliographystyle{prsty}

\title{Crystal defects and spin tunneling in single crystals of
Mn$_{12}$ clusters \vspace{-1mm} }

\author{J.M. Hernandez, F. Torres and J. Tejada}

\address{Dept. Fisica Fonamental, Univ. Barcelona. Diagonal 647. 08028 Barcelona.
Spain}

\author{E. Molins}

\address{Institut de Ci\`{e}ncia de Materials de Barcelona (CSIC).
Campus UAB. 08193 Cerdanyola. Spain\\ \smallskip {\rm(Received
\today)} }

 \maketitle

\begin{abstract}
The question addressed in this paper is that of the influence of
the density of dislocations on the spin tunneling in Mn$_{12}$
clusters. We have determined the variation in the mosaicity of the
crystal structure of fresh and thermally treated single crystals
of Mn$_{12}$ by analyzing the widening of low angle X-ray
diffraction peaks. It has also been well established from both
isothermal magnetization and relaxation experiments that there is
a broad distribution of tunneling rates which is shifted to higher
rates when increases the density of dislocations.
\end{abstract}
\begin{flushleft}
 \end{flushleft}

Within the last few years molecular clusters have emerged as a
truly interdisciplinary field. This is so because these materials
allow to test the border between quantum and classical mechanics
\cite{1}, they may be used as a hardware for quantum computers
\cite{4,5} and for low temperature magnetic cooling \cite{6}. The
magnetic hysteresis in molecular clusters results from the
existence of $2S+1$ spin levels in the two wells of the magnetic
anisotropy barrier. These spin levels correspond to the different
projections of the total spin of each molecule on its easy axis.
Mn$_{12}$ molecular clusters have $S=10$ at low temperature and
are equivalent to a single domain particle with constant modulus
of its magnetic moment, $20 \mu _B$, the orientation of which
depends on the ratio between the temperature and the barrier
height existing between the up and down orientations. The
occurrence of magnetic relaxation at temperatures at which the
thermal fluctuations die out is due to spin resonant tunneling
between degenerate $S_Z$ states in the two wells of the anisotropy
potential wells \cite{7,8,9,10,11,12,13,14,15,16,17}. To the first
approximation the spin Hamiltonian used previously
\cite{7,8,9,10,11,12,13,14,15,16,17} to fit the magnetic data
obtained for the different Mn$_{12}$ molecular clusters is written
as

\begin{equation} {\cal H}=-DS_Z^{2}+ {\cal H}'+ {\cal H}_{dip}+
{\cal H}_{hf} \label{eq1} \end{equation}

\noindent where $D= 0.65$ K  \cite{18,19,20} and ${\cal H}'$
contains anisotropy terms of fourth order of the spin operator
\cite{18,19,20,21} which depend on the symmetry of the crystal.
The two last terms correspond to the contribution of both dipolar
and hyperfine fields to the transverse magnetic field. The first
term of equation \ref{eq1} generates spin levels $S_Z$ inside each
well while the symmetry violating terms inducing tunneling are
those associated with the transverse component of the magnetic
anisotropy and the transverse dipolar and hyperfine fields. Very
recently, however, Chudnovsky and Garanin \cite{22} have suggested
that tunneling due to the magnetoelastic coupling, ${\cal
H}_{me}$, may be even larger that those due to the terms written
in equation \ref{eq1}. This may be so as a consequence of the
local transverse anisotropy and magnetic fields associated to
dislocations. Here we present experimental evidence on the effect
of defects, mostly dislocations, on both the rate of spin
tunneling and the law of relaxation in two Mn$_{12}$ single
crystals.

The Mn$_{12}$Ac forms a molecular crystal of tetragonal symmetry
with the lattice parameters $a =1.732$ nm and $c= 1.239$ nm
\cite{23}. The unit cell contains two Mn$_{12}$O$_{12}$ molecules
surrounded by four water molecules and two acetic acid molecules.
The Mn$_{12}$ 2Cl-benzoate, Mn$_{12}$Cl, forms a molecular crystal
of orthorhombic structure with the lattice parameters $a = 2.275$
nm, $b = 1.803$ nm and $c = 1.732$ nm with two molecules per unit
cell \cite{24,25}. The magnetic core and the local symmetry of
each molecule are identical to the case of Mn$_12$Ac with the only
difference that the easy axes of Mn$_{12}$Cl lie alternatively on
the direction ($011$) or ($0\overline{1}1$), being nearly
perpendicular to their nearest neighbors. Fresh single crystals of
Mn$_{12}$Ac and Mn$_{12}$Cl were first characterized by X-ray
diffraction techniques and then we carried out the magnetic
studies. The next step was to cycle the temperature of the single
crystals between 80 K and 300 K by introducing them in liquid
nitrogen during five minutes and in water for also five minutes.
This thermal cycle was repeated four times. Then we performed new
X-ray diffraction and magnetic characterization. A second heat
treatment, similar to the first one, followed by both X-ray and
magnetic characterization was also performed. During all these
operations the single crystals were kept glued on the top of a
glass capilar.

The thermal treatment suppose a rapid change in the temperature of
the surrounding air of the crystal of about 200 K. When the
crystal core is still at the initial temperature, the surface of
the crystal starts to cool down. This large temperature gradient
generates radial and tangential tensions that favor the
propagation of dislocations across the crystal, probably starting
at point defects frozen during the growing of the crystal and
during the X-ray irradiation (it should be taken into account that
for the X-ray characterization the crystals were irradiated during
24 hours). The extension of these dislocations by the whole
crystal converts an initial single crystal in a multidomain
crystal which each element is slightly misaligned in front of its
neighbors. This is what is known as a mosaic crystal and the
amount of misalignment is related to the widening of the
diffraction peaks. The peak widening observed in our experiments
is of some tenths of a degree. Due to this low value, the crystals
after the heat treatments are still considered as a single
crystals but with a larger mosaicity.

To better determine the variation in the mosaicity with the heat
treatment \cite{26,27,28}, we have focused our attention on low
angle reflection peaks in order to minimize the widening
associated to the lack of monochromaticity ($MoK_{\alpha1}$ and
$MoK_{\alpha2}$). After checking the crystal parameters and
getting the crystal orientation by using a four-circle
single-crystal X-ray diffractometer, we choose the reflections
($\pm 2$ $\pm 2$ $\pm 2$) for the comparison before and after the
heat treatment as they were low angle and intense.  In Figure 1 we
show the $\omega - \theta$ plot of the ($\overline{2}22$)
reflection, for the Mn$_{12}$Cl single crystal, before (left),
after the first heat treatment (center) and after the second heat
treatment (right). The insert clearly shows the enlargement of the
reflection peak along $\omega$ due to the increase of the
mosaicity while keeping constant the $2\theta$-width. In the final
state the flattening in $\omega$ even overcomes the scan width.
Assuming that the average distance between dislocations is
inversely related to the $\omega$-widening and taking into account
that such $\omega$ widening approximately doubles after each
thermal treatment, see Figure 1, it may be conclude that the
number of dislocation increases near an order of magnitude after
the heat treatment. Similar results were observed in the low angle
diffraction peaks for the Mn$_{12}$Ac single crystal.

The performed magnetic characterization of the single crystals
before and after the heat treatment, include hysteresis cycles at
different temperatures and relaxation experiments over the
resonance at zero field by changing the sweeping rate of the
magnetic field . The first experimental evidence that something is
going on with the heat treatment, came out from the magnetic
relaxation data at zero field, see Figure 2. In these measurements
the single crystals were first placed in a field $H= 30$ kOe at
$T=$ 10 K, then they  were cooled until the desired temperature
and the field was then switched off. The data displayed in Figure
2 and those obtained at $T= $2.0 and 2.2 K for the two single
crystals show that both the amount of magnetization relaxing per
unit time and the relaxation rate are larger after the heat
treatments. To see better this effect, we show in figure 2B the
calculation of the relaxation rate, $\Gamma $, at each time, which
was deduced from the relaxing data by using the differential
exponential law ($\Delta M = -M \Gamma \Delta t$). All these
relaxation curves are, however, pretty well fitted, in the entire
time interval, by stretched exponential functions, $M(t) =
M_{0}\exp(-a{\cdot}t^{b})$ being both $a$ and $b$ depending on the
concentration of dislocations.

In Figure 3 we show the field dependence of the differential
susceptibility $\partial M/\partial H$ deduced from the hysteresis
measurements recorded at two different temperatures for the
Mn$_{12}$Cl single crystal before and after the second heat
treatment. The position of the resonant peaks, where $dM/dH$ is
maximum, does not change with the heat treatment, indicating that
the internal structure of the Mn$_{12}$ clusters as well as the
spin levels are not influenced by defects. The sweeping rate of
the magnetic field in all these experiments was 10 Oe/s. In Table
I we give the values of the area under each peak $\partial
M/\partial H$ at the resonance fields at different temperatures.
These data show that the intensity and width of the peak at zero
field increase at all temperatures after the heat treatment. This
corresponds to have a larger amount of magnetization tunneling at
zero field at all temperatures. A similar enhancement is also
detected for the tunneling processes contributing to the second, 5
kOe, and third resonance, 10 kOe, when the hysteresis is recorded
at low temperature. The reduction in the intensity of the peaks at
larger fields and higher temperature after the heat treatment with
respect to those observed with the fresh material, corresponds to
the fact that as we are studying the demagnetization process there
is less magnetization to tunnel after crossing the zero field
resonance. The temperature tends to reduce the differences in the
observed $\partial M/\partial H$ peaks at high fields for the
fresh and treated crystals as it should be due to the variation of
the equilibrium magnetization with temperature and to the fact
that the detected tunneling rates at large times do not change to
much with the heat treatment. That is, there is always a fraction
of molecules not affected by defects. These molecules show the
same relaxation rate before and after the heat treatment.

Let us start the interpretation by commenting on the data shown in
Figure 2B. These data strongly support that the spin relaxation of
our sample is a superposition of exponential decays with different
rates

\begin{equation}
 M(t) = M_0 \sum_i \exp (-\Gamma _i t)
\end{equation}

Where $M_0$ is the initial magnetization, that is at $t=$0, and
$\Gamma _i$ should correspond to the relaxation rate for the
different molecules affected or not by dislocations. The molecules
near the core of the dislocations are those for which the
magnetoelastic coupling is the largest and, consequently, show the
strongest tunneling effect. That is, these are the molecules owing
the largest tunneling rates. As we move away from the nucleus of
the dislocations both the transverse anisotropy and transverse
field become smaller and consequently decreases the spin tunneling
rate for the molecules located far of dislocations, increasing
therefore the number of molecules affected by dislocations and
their tunneling rates. This is the reason why there is a
distribution of relaxation times and the measured tunneling rates
decreases as time is running. That is, the fresh single crystals
and those thermally treated show a broad distribution of tunneling
splittings, $\Delta $. In fact, this is the reason why the
relaxation rate of the magnetization for the fresh crystals show
also a time-dependence. The gravity center of this distribution
shifts to larger tunneling rates after the heat treatment as a
consequence of having a larger density of dislocations. This
interpretation is also supported by the results obtained from the
hysteresis measurements. The crystals after the heat treatments
have higher density of dislocations and show, at all temperatures,
larger tunneling magnetization at the zero field resonance. The
fact that the amount of magnetization relaxing at the resonance
located at high fields decreases after the heat treatment reflects
the fact that there is less magnetization to relax.  We may,
therefore, conclude that the variation in the crystal mosaicity
detected in the X-ray experiments is fully correlated with the
increase in the number of molecules relaxing faster which, as a
matter of fact, are the molecules affected by dislocations.

We have also found that when sweeping the field through the zero
field resonance, the relaxing magnetization normalized to its
initial value, scales as a function of $T\ln(\nu \: r)$. This
reflects the fact that in the presence of dislocations occur both,
a broad distribution and the overlapping of the tunneling
probability for the different spin levels.

Summarizing, we have shown and discussed magnetic data for two
molecular single crystals which indicate that dislocations modify
substantially the spin tunneling rate. The magnetic relaxation law
is strongly affected by the existence of defects as a consequence
of the overlapping of tunneling splittings for the different spin
levels. The results are in agreement with a recent work on the
nature of spin tunneling in Mn$_{12}$-Acetate \cite{29} and in the
broading of the EPR linewidths due to dislocations \cite{30}.

Acknowledgments.- The authors thank the EC Grant number
IST-1999-29110 for financial support.

\begin{table}
\begin{tabbing}
   (Ksssd) \= Tdddreatmnt \= 0ddkOe \= 5kddOe \= 10kddOe \= 15kddOe\kill
   \>  {\bf Heat} \>  {\bf Resonance's area ($\times 10^{-4}$ emu)} \>  \>  \>   \\
  $\bf T$ (K) \>  {\bf treatment} \> {\bf 0 kOe} \> {\bf 5 kOe} \> {\bf 10 kOe} \> {\bf 15 kOe }\\
  1.8  \> Before \> 19.6 \> $< 1$ \> 10.0 \> 54.6 \\
    \> After 1st \> 25.5 \> $< 1$ \> 14.1 \> 54.7 \\
    \> After 2nd \> 28.8 \> $< 1$ \> 12.8 \> 47.7 \\
  2.0  \> Before \> 17.3 \> 5.2 \> 33.6 \> 46.5 \\
    \> After 1st \> 20.6 \> 7.9 \> 44.1 \> 40.4 \\
    \> After 2nd \> 31.4 \> 10.1 \> 26.6 \> 31.7 \\
  2.2  \> Before \> 22.1 \> 15.5 \> 60.9 \> 24.4 \\
    \> After 1st \> 33.5 \> 24.0 \> 52.6 \> 23.3 \\
    \> After 2nd \> 33.2 \> 25.0 \> 51.6 \> 16.8 \\
  2.4  \> Before \> 22.5 \> 44.6 \> 43.8 \> 4.1 \\
    \> After 1st \> 33.6 \> 44.1 \> 28.3 \> 3.2 \\
    \> After 2nd \> 38.1 \> 46.0 \> 29.3 \> 2.4 \\
\end{tabbing}
 \caption{Areas below the differential susceptibility
$\partial M/\partial H$ curves at the resonance peaks for the
Mn$_{12}$Cl single crystal before and after the first and second
heat treatment deduced from hysteresis measurements recorded at
different temperatures.} \end{table}

\begin{figure}
 \centering
 \epsfig{width=4.5in,file=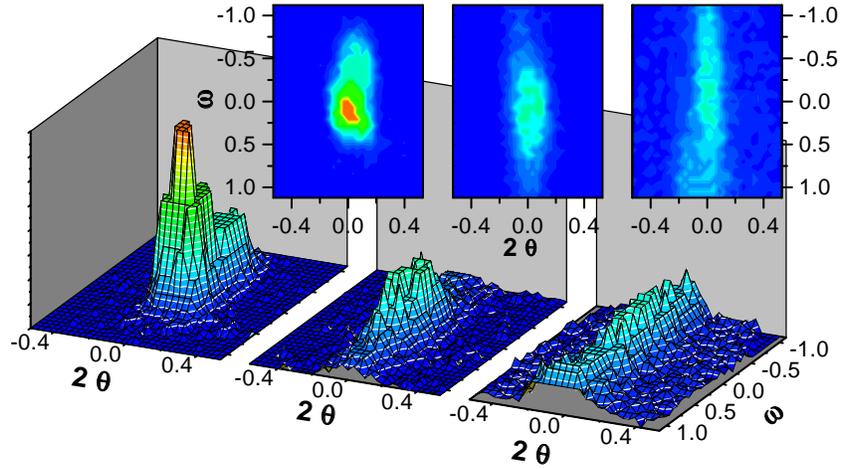}
 \caption{  $\omega -\theta$ plot of the ($\overline{2}22$)
reflection of the Mn$_{12}$Cl crystal: before (left),  after one
heat treatment (center) and after two heat treatments (right) .
The inserts clearly show the enlargement of the peak along
$\omega$ due to the increase in mosaicity after the heat
treatments, while keeping a constant $\theta$-width. The
flattening in $\omega$ even overcomes the scan width.}
\end{figure}

\begin{figure}
\centering \epsfig{width=4.5in,file=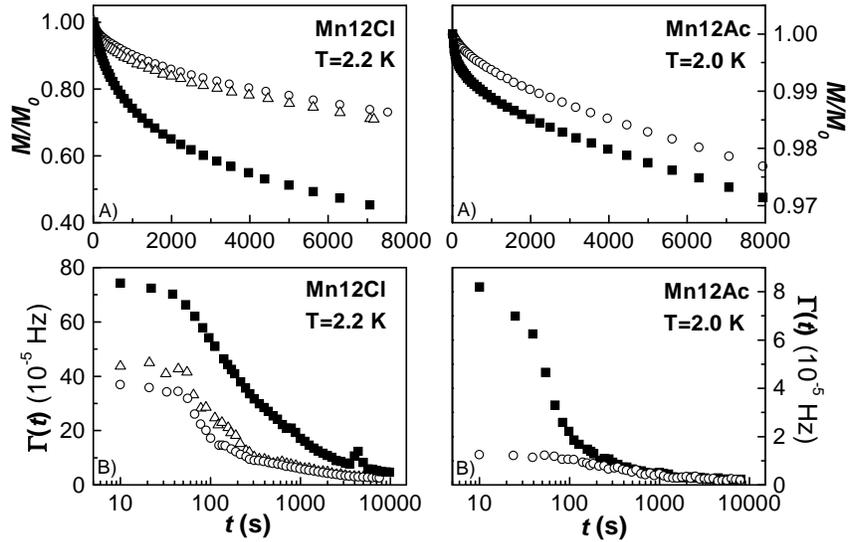} \caption{  A)
Magnetic relaxation of single crystals of Mn$_{12}$Ac and
Mn$_{12}$Cl at different temperatures before (open circles) and
after (open triangles, solid squares) the heat treatments.
B)Evolution with time of the effective relaxation rate
corresponding to the data depicted on A. The relaxation rates at
each time have been calculated as the time derivatives of the
logarithm of the magnetization. }
  \label{fig:2}
\end{figure}

\begin{figure}
\centering \epsfig{width=3in,file=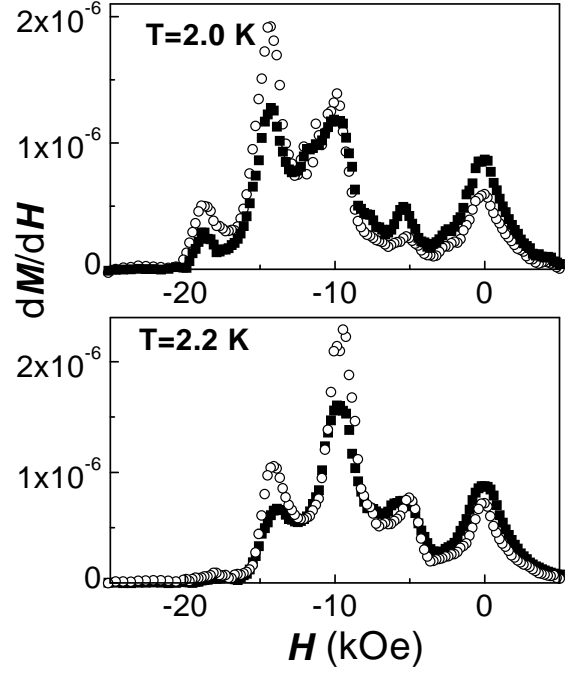} \caption{
Differential susceptibility, $\partial M/\partial H$, at the
resonance fields deduced from the hysteresis cycles at different
temperatures for the Mn$_{12}$Cl single crystal, before (open
circles) and after (solid squares) the heat treatments.}
  \label{fig:3}
\end{figure}

\end{document}